# Satellite-derived solar radiation for intra-hour and intra-day applications: Biases and uncertainties by season and altitude

Carpentieri, A.[a,b], D. Folini[b], M. Wild[b], L. Vuilleumier[c], A. Meyer[a]

a Bern University of Applied Sciences, Quellgasse 21, 2501 Biel, Switzerland
b Institute for Atmospheric and Climate Science, ETH Zurich, Universitaetstrasse 16, 8092 Zurich, Switzerland
c Federal Office of Meteorology and Climatology MeteoSwiss, Chemin de l'Aérologie 1, 1530 Payerne, Switzerland
Corresponding author: alberto.carpentieri@bfh.ch

**Abstract.** Accurate estimates of the surface solar radiation (SSR) are a prerequisite for intra-day forecasts of solar resources and photovoltaic power generation. Intra-day SSR forecasts are of interest to power traders and to operators of solar plants and power grids who seek to optimize their revenues and maintain the grid stability by matching power supply and demand. Our study analyzes systematic biases and the uncertainty of SSR estimates derived from Meteosat with the SARAH-2 and HelioMont algorithms at intra-hour and intra-day time scales. The satellite SSR estimates are analyzed based on 136 ground stations across altitudes from 200 m to 3570 m Switzerland in 2018. We find major biases and uncertainties in the instantaneous, hourly and daily-mean SSR. In peak daytime periods, the instantaneous satellite SSR deviates from the ground-measured SSR by a mean absolute deviation (MAD) of 110.4 and 99.6 W/m$^2$ for SARAH-2 and HelioMont, respectively. For the daytime SSR, the instantaneous, hourly and daily-mean MADs amount to 91.7, 81.1, 50.8 and 82.5, 66.7, 42.9 W/m$^2$ for SARAH-2 and HelioMont, respectively. Further, the SARAH-2 instantaneous SSR drastically underestimates the solar resources at altitudes above 1000 m in the winter half year. A possible explanation in line with the seasonality of the bias is that snow cover may be misinterpreted as clouds at higher altitudes.

**Key points**
- We assess intra-day/-hour solar radiation derived with SARAH-2 and HelioMont
- Satellite estimates compared with 136 pyranometers at altitudes from 200 to 3570 m
- Daytime intra-hour MAD amounts to 91.7 W/m2 (SARAH-2) and 82.5 W/m2 (HelioMont)
- SARAH-2 drastically underestimates the SSR at high altitudes in winter
- The biases depend on season and altitude, with snow cover as a potential challenge

# 1. Introduction

Accurate estimates of the solar resources at a specific time and location are a critical requirement for a variety of applications. On time scales of decades to millennia, solar resource estimates are relevant for studying climatological conditions and to foster our understanding of planetary processes and climatological trends (e.g., Wild, 2009). On shorter time horizons, accurate solar resource estimates are needed for assessing the profitability of photovoltaic development projects (e.g, Suri et al., 2011; Gueymard, 2012, 2014; Kleissl, 2013; Vuilleumier et al., 2020) but also for power and energy system simulations. At time scales of minutes to days, accurate solar resource estimates are beneficial for grid balancing applications in power systems with a high share of photovoltaic energy and for reliable short-term forecasts of solar resources. Accurate forecasts play an important role in renewable power production planning, grid integration, and energy trading applications. These applications become more and more important as the installed photovoltaic power capacity is expected to continue its expansion around the world in the next decades (Bojek et al., 2021; Kuhn et al., 2018).

Surface solar resources can be measured by meteorological ground stations and they can also be estimated from satellite remote sensing of the Earth. Geostationary satellites such as the Meteosat family provide visible and infrared images from which the surface solar resources can be estimated (Cano et al., 1986). Satellite-derived estimates of the solar resources are particularly valuable in regions for which high-quality ground measurements are infeasible or unavailable.

This study investigates uncertainties and systematic biases in satellite-derived estimates of surface solar radiation (SSR) under all sky conditions. The SSR is also called shortwave downward radiation, global horizontal irradiance, surface incoming shortwave radiation, and global solar radiation in different related research contexts. The SSR is the total solar energy flux incident on a horizontal plane from the sky half dome typically between roughly 300-2800 nm. We focus on SSR maps from the geostationary Meteosat satellites, which cover a wide field of view from latitudes of about 65°N to 65°S. In contrast, ground station pyranometers usually provide SSR measurements that are significantly more precise but their spatial representativeness is limited. Disregarding complex topographies, monthly SSR in situ observations can be considered representative of a 1°x1° area (Schwarz et al., 2018). However, higher time resolutions reduce the representativeness of the ground stations (e.g., Li et al., 2005). For hourly observations, the ground stations can be reliable for an area of only a few kilometers. In this study, we investigate the accuracy and biases of SSR estimates at high temporal resolution with a view towards intra-day and intra-hour applications. We consider the bias to be the difference between the satellite-estimated and the ground-measured SSR.

In this study, we investigate the SSR from two different retrieval products – SARAH-2 and HelioMont – derived from the Spinning Enhanced Visible and InfraRed Imager (SEVIRI) instrument onboard the Meteosat satellite. Previous studies examining different Meteosat retrieval products found that the uncertainty of the monthly-mean SSR of the SARAH data product, a predecessor of SARAH-2, has a similar magnitude as the measurement uncertainty of ground-based pyranometer networks. The studies focused on monthly-mean analyses with a view towards climatological applications. Specifically, Müller et al., 2015 validated the SSR of SARAH with pyranometer measurements of the Baseline Surface Radiation Network (BSRN) and the Global Energy and Balance Archive (GEBA). The authors found that the mean absolute deviation (MAD) of the monthly-mean SSR amounted to 5.5 W/m$^2$. Riihelä et al., 2015 studied the accuracy of the SARAH SSR with ground station measurements from Scandinavia. They found that the SARAH algorithm was overall able to capture the monthly and daily-mean SSR and its seasonality but that it also exhibits significant negative biases. The authors found the monthly-mean and daily-mean root mean square deviation (RMSD) across all stations to be 8.3 and 17.0 W/m$^2$, respectively. Mazorra Aguiar et al., 2019 investigated hourly SARAH SSR estimates for the Canary Islands in the years 2010-2011. They reported relative RMSDs in the hourly-mean SSR of about 40% when comparing to 22 meteorological ground stations on the islands. Moreover, Castelli et al., 2014 found mean absolute deviations in the HelioMont hourly-mean SSR of 40 and 52 W/m$^2$ at the ground stations of Payerne and Davos, Switzerland. When studying systematic under- and overestimations, they reported mean biases (MBD) of 2 and -6 W/m$^2$ in hourly-mean SSR at these stations. Pfeifroth et al., 2018 investigated trends and variability in the SSR across Europe using the SARAH-2 product and found monthly mean MADs of 6.9 W/m$^2$. Babar et al., 2019 analyzed the SARAH-2 dataset at 31 meteorological ground stations across Norway. They found MAD values in the monthly- and daily-mean SSR of 5 and 11.8 W/m$^2$, respectively, in SARAH-2. Greuell et al., 2013 introduced and characterized a Meteosat based SSR data product that they validated by exploiting eight stations of the BSRN network for the year 2006. On average, the dataset showed an MAD of 7 W/m$^2$ and an hourly RMSD of 65 W/m$^2$. The median value of the hourly all-sky SSR RMSD, taken across the eight stations, was 75 W/m$^2$. The same SSR data product was studied by Dirksen et al., 2017. They considered daily values over a period of 12 years. The authors validated the SSR product by means of a ground station network with 32 stations in the Netherlands. They detected a seasonality in the biases that was positive in summer (2.7 W/m$^2$) and absent in winter. All previous studies focused on the accuracy of time-aggregated SSR, mostly of the monthly- and daily-mean SSR. Surprisingly, the accuracy and biases of the intra-hour solar resource satellite estimates and, in particular, the instantaneous SSR estimates derived from single Meteosat SEVIRI images have not been investigated yet. However, these maps are the most relevant ones for intra-hour and intra-day applications such as near-real-time estimates of the photovoltaic (PV) production in a specific region, and for short-term forecasting the PV production at lead times of minutes and hours. Such short-term applications have motivated this work.

The objective of our study is to address that research gap by characterizing the accuracy and biases of the SSR from Meteosat SEVIRI on intra-hour and intra-day time scales.

The previous studies included SSR estimates from both day- and nighttime periods. This clearly results in lower uncertainties and biases than daytime-only analyses. Our study is motivated by solar PV applications. Nighttime is of less interest in this study, therefore, so we exclude nighttime from our analysis. Including nighttime periods would result in SSR accuracies that would be misleadingly low for the case of PV applications. Moreover, the previous studies were performed for regions at relatively narrow altitude ranges, for example, the Dutch lowlands and inland regions of Norway. In contrast, ours is the first SSR study to cover approximately the full range of altitudes at which solar resource assessments are needed and at which PV systems operate worldwide. Specifically, we investigate the accuracy of satellite-derived SSR estimates at altitudes from 200 m to 3570 m above sea level. Regions at higher altitudes including mountainous areas offer high SSR resources even in boreal winter. These resources can serve to compensate the reduced PV production in the winter season (Kahl et al., 2019). It is therefore important to accurately characterize the solar resource potential also at higher altitudes. Seasonal changes in land surface cover and albedo can render the SSR estimation more challenging in these regions.

Our study is applied to Switzerland. This central European region offers diverse land surface conditions and surface albedos, covers a wide range of altitudes and seasonal conditions, and provides a dense and well-maintained ground station network including a BSRN station. Our findings are not specific to Switzerland in that they can be applied to other regions with comparable atmospheric and surface conditions.

This study is structured as follows. Section 2 describes the radiation data by summarizing the satellite instrument and the derived data products. Section 3 gives an overview of the ground instruments and the radiation data measured by the ground stations. Section 4 reports the methods, data processing and the metrics applied. Our results are discussed in section 5. Finally, section 6 presents the conclusions of our study.

**2. Satellite estimates of surface solar radiation**

The satellite-derived SSR datasets assessed in this study are the SARAH-2 and the HelioMont data products (Müller et al., 2015; Stöckli, 2013). Heliosat-type data products, such as SARAH-2 and HelioMont, are derived based on multiple sources of data. This included SEVIRI observations of clouds but also datasets on other radiatively-relevant atmospheric constituents, notably aerosol and trace gases. The observations of the SEVIRI instrument onboard the geostationary Meteosat-11 satellite enable to infer cloud effects, whereas aerosol, water vapor and ozone column data are used to compute clear-sky

SSR estimates which are then modified by the cloud effect derived from SEVIRI. The HelioMont and SARAH-2 maps come at spatial resolutions of 0.02°x0.02° and 0.05°x0.05°, respectively. We make use of the highest available temporal resolution of 15-minutes instantaneous SSR maps from HelioMont and 30-minutes instantaneous SSR maps from SARAH-2 in 2018. The SSR estimates are compared with the pyranometer ground station measurements by matching each ground station with the corresponding pixel of the satellite-derived SSR at the closest matching SEVIRI scan time. Only daytime values of SSR are included, identified via the solar zenith angle (SZA) and in free horizon situation, which varies for every different location. Only daytime timestamps are considered in the analysis by requiring the sun to be above the horizon using location-dependent SZA-threshold selection criteria, as detailed in section 3.

**2.1 Meteosat SEVIRI instrument**
The SEVIRI instrument onboard the Meteosat Second Generation satellites continuously images the Earth. SEVIRI's primary purpose is to provide thermal and visible images of the Earth for numerical weather forecasting. The SEVIRI instrument has twelve channels in the visible and infrared spectral range. The temperatures of clouds and of Earth's surface can be derived based on the infrared channels. The geostationary Meteosat orbit enables high-frequency scan cycles of only 30 ms duration per East-to-West scan line, and a dissemination of the resulting instantaneous scan images every 15 minutes (Müller et al., 2010). The SEVIRI field of view ranges from -65° to +65° in latitude and longitude. The SSR analyzed in this study is estimated from scans of the Meteosat-11 satellite positioned at 0°N, 0°E nadir where it has been commissioned in 2015.

**2.2 SARAH-2 surface solar radiation**
Our study makes use of the SSR estimates derived with the SARAH-2 method (Müller et al., 2015; Pfeifroth et al., 2017; EUMETSAT, 2022). SARAH-2 computes surface solar radiation and its components including direct and diffuse radiation. The dataset is provided by the EUMETSAT Satellite Application Facility on Climate Monitoring (CMSAF). SARAH-2 is a multi-decade radiation climate record covering the SEVIRI field of view. The SARAH-2 algorithm estimates the SSR from Meteosat SEVIRI scans every 30 minutes at a resolution of approximately 5x5 $km^2$ at the ground. Those are instantaneous SSR estimates, not temporal averages over 30 minutes. The instantaneous SARAH-2 SSR is used at 30-minute intervals for the year 2018. The year 2018 is selected for the present study to enable maximum overlap with the other available data sources, specifically the HelioMont-derived SSR, the SwissMetNet ground station measurements and the BSRN station data described below.
The SSR calculation of the SARAH-2 algorithm computes the effective broadband albedo of clouds from the SEVIRI channels. This calculation is performed based on the Heliosat method (Moeser et al., 1984; Cano et al., 1986; Müller et al., 2009) that compares the measured reflectances from SEVIRI irradiance maps to clear-sky top-of-atmosphere reflectances of the Earth's surface and atmosphere to derive the cloud radiative effects. A radiative transfer model correction is performed to account for the spectral impact of clouds on the SSR. A look-up table approach is pursued to speed up the application of the libRadtran radiative transfer model (Mayer et al., 2015). The SSR is computed in a spectrally resolved manner based on the correlated-k method (Kato et al., 1999). The radiative transfer calculation and look up of the spectrally resolved irradiances take into account the radiative scattering and absorption of atmospheric aerosols, water vapor and ozone. The SSR reported in SARAH-2 is based on the sum of the spectral contributions to the radiative flux, also known as broadband flux (Müller et al., 2015). There are approximately 2% of missing values in the SARAH-2 SSR dataset, which are not considered in the analysis.
Daily water vapor column values are adopted from the operational daily ECMWF model analysis at 12 UTC. Monthly-mean ozone is assumed in the SARAH-2 SSR estimation from ERA Interim (Dee et al., 2011). The monthly mean aerosol properties are based on the climatology of the Monitoring Atmospheric Composition and Climate (MACC) project (Inness et al., 2013; Müller et al., 2015b).

## 2.3 HelioMont surface solar radiation

An additional Meteosat-based SSR dataset analyzed in this study is the HelioMont SSR (Stöckli, 2013). HelioMont is an algorithm from the Heliosat family like SARAH-2. Similar to SARAH-2, HelioMont computes the radiative forcing of clouds from the SEVIRI infrared and visible channels. The clear-sky solar energy fluxes cannot be observed but are computed by a radiative transfer model without using any satellite observations. The clear-sky irradiance is obtained from look-up tables that have been constructed from the libRadtran radiative transfer model. Like in the SARAH-2 algorithm, the tables account for the radiative impacts of atmospheric aerosol, water vapor and ozone on the surface solar radiation. Like in all Heliosat-type methods, the all-sky surface solar energy fluxes in HelioMont and SARAH-2 are obtained by combining the estimate of the clear-sky radiative flux with the inferred cloud effect. In HelioMont, the all-sky SSR is obtained as the product of a clear-sky index and the clear-sky SSR derived from a radiative transfer model look-up table. The assumed aerosol properties are in accordance with the CAMS aerosol dataset (Inness et al., 2019) whereas the water vapor and ozone are taken from ERA Interim (Dee et al., 2011). The aerosol, water vapor and ozone datasets come at a 6-hour temporal and 0.5° spatial resolution. A characteristic feature of the HelioMont algorithm is that, unlike SARAH-2, it was designed to respond to fast changes in the Earth's surface albedo, such as from fresh snow cover. HelioMont discriminates clouds and snow by relying on the fact that ground albedo evolves slowly whereas cloud albedo evolves rapidly. To enable the distinction of cloud and surface albedos, HelioMont computes the clear-sky albedo over the course of a day based on the surface albedo and cloud mask of the previous days. To this end, clear-sky observations of the visible and infrared channels are collected from the past up to ten days to compile a clear-sky albedo map characterizing the ground albedo. The observations are weighted by the time that passed since they were made and by the confidence in the clear-sky state. Missing observations are filled based on models of the expected reflectance and brightness temperatures (Stöckli, 2013). HelioMont makes use of the SEVIRI high-resolution visible channel (0.45-1.1 µm) and multiple infrared channels to this end. Subsequently, it compares the all-sky reflectances measured by the SEVIRI imager with the estimated clear-sky reflectances and derives a cloud and a clear-sky index.

The instantaneous HelioMont SSR estimates are provided every 15 minutes at 0.02°x0.02° resolution for the year 2018 over the SEVIRI field of view. The fraction of missing values in the HelioMont SSR is 0.24% in the studied time period. Missing values have been removed from the analysis.

## 3. Ground measurements

The SSR pyranometer measurements analyzed in this study originate from two ground networks: The meteorological monitoring network SwissMetNet (SMN) of the Swiss national weather service MeteoSwiss and the Swiss Alpine Climate Radiation Monitoring (SACRaM) network. The SCARaM network is dedicated to providing high-accuracy radiation measurements obtained by following the guidelines of the BSRN. The locations and altitudes of all ground stations are provided in Table A1 (supplementary material). The SMN and SACRaM stations are distributed densely across all regions of Switzerland. In particular, they cover a wide range of altitudes from 200 m at the Lake Maggiore to 3570 m above sea level at the Alpine research station Jungfraujoch. An impression of the geographical distribution of the stations may be obtained from Figure 1. All ground stations are operated and maintained in accordance with the high quality standards defined by the World Meteorological Organization with regard to measurement devices, their location, operation, maintenance and quality control (WMO, 2018). All measurement data have been quality controlled in accordance with these standards.

## 3.1 SwissMetNet

Our study employs pyranometer measurements from 133 meteorological ground stations of the SwissMetNet distributed across Switzerland (SwissMetNet, 2022). The SMN is an automatic monitoring network operated by the Federal Office of Meteorology and Climatology MeteoSwiss as Switzerland's national weather service. The SMN is one of the densest meteorological ground station networks around the world. SSR is measured at 133 of its 160 stations which are equipped with well-maintained environmental sensing systems. Most of the SMN ground stations are equipped with high-precision pyranometers of type Kipp & Zonen CM21 and CMP21 which provide SSR integrated over 285-2800 nm as measured at the respective station. This study analyzes data of the year 2018 provided with a 10-minute time step. The SMN SSR measurements are 10-minute averages of SSR measurements at an original 1-Hz sampling rate of the pyranometers. Time stamps with missing SSR values are removed from the dataset and are not considered in the analysis. No gap filling is used. Across all 133 stations, the fraction of missing values ranges from 0% to 4%. The mean fraction of missing values across all stations and time steps is 0.2% in the study time period.

### 3.2 BSRN
The Baseline Surface Radiation Network (BSRN) is a global surface radiation monitoring network that aims to detect possible changes and trends in the surface radiation budget of the Earth and radiative impacts of climate change (Ohmura et al., 1998). This study makes use of SSR measurements from the BSRN station in Payerne, Switzerland, situated at 46.82°N, 6.94°E at an altitude of 491 m above sea level. The BSRN station Payerne follows the BSRN operation guidelines (McArthur, 2005). It is also one of the SACRaM stations and measures SSR with a high-precision pyranometer of type Kipp & Zonen CM22. The SACRaM and the SMN pyranometers employ thermopiles whose black surfaces warm up from the incident shortwave solar radiation coming from the upper hemisphere. The associated thermal energy is converted into electrical energy which results in a thermopile voltage from which the SSR is computed. Vuilleumier et al., 2014 discussed sources of measurement errors for the BSRN pyranometer and quantified the associated measurement uncertainties to within 1.8% of the measured values. They demonstrated that the Payerne BSRN pyranometer measurements are in line with BSRN's high quality targets.

### 3.3 SACRaM network
The Swiss Alpine Climate Radiation Monitoring (SACRaM) network provides long-term high-precision measurements of SSR for climate monitoring objectives in Switzerland. In addition to the BSRN station in Payerne, data from two other SACRaM stations are used here which are located in Locarno at Lake Maggiore and in the Swiss Alps at Davos. The SACRaM stations are also operated and maintained in accordance with BSRN standards and, thus, also provide SSR measurements at BSRN accuracy. The SACRaM data are 1-minute averages of SSR measurements at 1-Hz sampling rate of the pyranometers. Relevant for this study is the quasi-daily maintenance during workdays, including cleaning, monthly calibration checks using on-site open absolute cavity radiometers, secondary standards traceable to World Radiometric Reference and thermal offset correction. SSR is measured with Kipp & Zonen CM21, CMP21, CM22 or CMP22 pyranometers depending on the station and time period.

### 4. Methods
The satellite-derived SSR estimates and the SSR ground measurements underwent different spatial and / or temporal averaging. The SSR satellite estimates constitute instantaneous maps of the surface solar radiation at scan time without any additional temporal averaging. In contrast, the ground-measured SSR is provided as 10-minutes average values. Thus, 10-minutes averages of in-situ point measurements from pyranometers are compared to the corresponding instantaneously captured SSR pixel representing a 0.02˚x0.02˚ or 0.05˚x0.05˚ surface area in the HelioMont and SARAH-2 products, respectively. The time series are matched by comparing the instantaneous satellite-derived SSR with the 10-minutes average value based on the satellite timestamp. Then, we define the SSR bias $y_{i,t}$ as the difference (or residual) between the satellite-derived SSR, $SSR^{\text{sat}}$, and the associated SSR measurement at the corresponding ground station, $SSR^{\text{ground}}$,

$$y_{i,t} = SSR^{\text{sat}}_{i,t} - SSR^{\text{ground}}_{i,t}$$

where $t$ and $i$ indicate the time dependence and the ground station, respectively. The following metrics are used to characterize the accuracy and the biases of the satellite-derived SSR estimates in this study:

$$RMSD = \sqrt{\frac{1}{n}\Sigma_{i,t} y^2_{i,t}}$$

$$MBD = \frac{1}{n}\Sigma_{i,t} y_{i,t}$$

$$MAD = \frac{1}{n}\Sigma_{i,t} |y_{i,t}|$$

$$rMAD = \frac{1}{n}\Sigma_{i,t} \frac{|y_{i,t}|}{SSR^{\text{ground}}_{i,t}}$$

where $RMSD$, $MBD$, $MAD$, and $rMAD$ denote the root mean square deviation (RMSD), mean bias deviation (MBD), mean absolute deviation (MAD) and the relative mean absolute deviation (rMAD), respectively. Furthermore, $n$ is the number of instances available for the metric computation for a single site. The MBD measures the mean bias whereas RMSD, MAD, and rMAD quantify the overall accuracy of the satellite-derived SSR estimates, including any potential biases. Thus, the MBD is the overall long-term difference between the estimated and the observed SSR. It quantifies the mean deviation of the SSR residuals from zero, and depends on the considered sites but also on the time of year, as discussed below. The RMSD, MAD and rMAD quantify the accuracy in terms of the scatter of the satellite SSR estimates around the in-situ SSR. The accuracy is affected by the MBD because the satellite SSR estimates are not bias-corrected in our study. The MBD can be corrected, for example, by computing site- and time-dependent long-term means with regard to in-situ SSR reference measurements, and subtracting those means. Such simple bias correction approaches may induce additional bias though. Another approach towards bias correction is based on addressing and removing the causes of the biases. For example, if systematic biases originate from issues of distinguishing clouds and snow cover, then the resulting biases may be corrected by a regression model with information related to cloud cover or snow cover as one of the regressors. The accuracy of the SSR satellite estimates in our study may be improved by correcting systematic biases. Other sources of inaccuracy, such as measurement noise, spatial or temporal sampling errors, can usually not be corrected.

We first compute the metric, e.g., the RMSD, at each ground station site over the year 2018 and then average the metric over all sites. The resulting metrics are presented in Tables 1–3. SSR monitoring and short-term forecasting are of interest only in daytime periods for the applications addressed by this study. Therefore, our study focuses on daytime periods and excludes nighttime from the analysis. Daily and monthly mean SSR values are calculated directly from the instantaneous and 10-minutes average SSR values in the respective day and month rather than from aggregated intermediate quantities such as the hourly mean or the daily mean SSR. Moreover, we only consider SSR measurements from SMN stations at times when the sun is above the local horizon to avoid shadowing effects from mountains. To this end, we apply station-specific thresholds, $SZA_i^{\max}$, on the maximum SZA that are computed based on horizon profiles obtained from a digital elevation model of MeteoSwiss. The model is used to compute the horizon for each SMN station. It determines the SZA and whether the sun was above or below the local horizon for each station and for all time steps of the year 2018. This enables us to compute the maximum elevation at which the horizon blocks the direct solar beam at the respective station and time. A single station-specific $SZA_i^{\max}$ is set for all seasons to simplify the processing. This results in a partial loss of the winter season SSR observations at stations with particularly high horizons. Less than half of the stations are affected by a low SZA threshold, so a sufficient number of stations contributes SSR ground measurements across all altitudes throughout the year. Further, we also require that a minimum number of measurements must be available to compute time-aggregated SSR values, specifically 2, 8, and 90 instances for hourly, daily, and monthly means, respectively. If fewer values are present, then the data for the corresponding time period and station are discarded from the analysis.

## 5. Results and discussion

We start with an aggregated view, quantifying SSR accuracy via comparing averages over the different ground stations. The analysis highlights the impact of temporal scales – instantaneous, hourly, daily, or monthly means – and gives a first impression of the performance of the two satellite products examined, SARAH-2 and HelioMont. We then illustrate that the differences found exceed potential contributions from uncertainties of in situ measurement and timing issues (satellite data are instantaneous, while in situ data are averages over several minutes). Finally, we take a more site-specific perspective, which highlights the crucial role of altitude for the accuracy and bias of satellite-based SSR, as well as the overall spatial and seasonal heterogeneity of the bias and accuracy pattern.

### 5.1 Site-averaged accuracy

Table 1 shows the accuracy of the daytime SSR for SARAH-2 and HelioMont whereas Table 2 presents these metrics over the entire day- and nighttime periods (i.e., daily means). Focusing on only the daytime periods strongly increases the bias magnitudes as can be seen from Tables 1 and 2. We also investigate the accuracy of the instantaneous SSR estimates across all SMN stations and the effect of time aggregation on the hourly, daily and monthly-mean SSR. The RMSD in the daytime SSR increases by a factor of 3-4 when going from monthly daytime mean to instantaneous SSR estimates. This factor is of similar magnitude as the ratios obtained by Li et al., 2005.

Both satellite products tend to underestimate the all-sky SSR with a systematic negative mean bias (MBD) on the order of -15 W/m$^2$ and -7 W/m$^2$ in the SARAH-2 SSR and the HelioMont SSR, respectively, when computed over all sites and the entire year 2018. The MBD detects the sign of systematic biases, so negative MBDs point to a systematic underestimation. The negative biases of the satellite-derived SSRs in both products cannot be explained by potentially suboptimal cleaning effects of the SMN pyranometers. Soiling reduces the ground-measured SSR. Soiling-induced biases would therefore increase the magnitude of the negative satellite SSR MBD. Thus, the actual negative satellite SSR biases might be even larger in magnitude than -15 and -7 W/m$^2$, respectively.

### 5.2 Potential contribution of in situ SSR measurement uncertainty

We quantify the uncertainty in the SSR measurements of the ground stations to carefully assess the accuracy of the satellite-estimated SSR. The measurement uncertainty of the three BSRN and SACRaM station pyranometers has been estimated to be at most 18 W/m$^2$ at full SSR signal of 1000 W/m$^2$ and significantly lower in case of smaller SSR values (Vuilleumier et al., 2014). The pyranometers at the BSRN/SACRaM station in Payerne and the SACRaM stations in Davos and Locarno get cleaned daily on workdays and undergo quality control in accordance with the BSRN standards. At the SMN stations, an SSR measurement uncertainty of at most 18 W/m$^2$ at full signal can in principle not be guaranteed. This is because the SMN pyranometers are not cleaned on a daily basis, as would be required by BSRN guidelines, but rather every two to three weeks. Moreover, the SMN pyranometers do not undergo a correction of potential thermal infrared loss offsets. Nor can their measurements be checked for deviations of directly-measured SSR in comparison to the sum of the diffuse and direct beam components because instruments for measuring the direct and diffuse radiation components are not available at SMN stations. Based on Vuilleumier et al., 2014, we conservatively estimate that these conditions amount to an overall maximum uncertainty (RMSD) of 25 W/m$^2$ of the SSR measurements at the SMN stations. This upper limit applies for instantaneous peak (noontime) SSR values, whereas we include also non-peak SSR values in our analysis. Non-peak SSR measurements should be associated with an uncertainty that is significantly lower than 25 W/m$^2$ (Vuilleumier et al., 2014).

| **Daytime SSR** | **Instantaneous** | **Hourly mean** | **Daytime mean** | **Monthly daytime mean** |
|---|---|---|---|---|
| RMSD SARAH-2 | 136.9 (133.7) | 116.7 (115.3) | 70.2 (73.6) | 41.7 (47.0) |
| MBD SARAH-2 | -15.0 (-26.2) | -15.7 (-27.4) | -13.8 (-29.3) | -15.6 (-29.9) |
| MAD SARAH-2 | 91.7 (87.7) | 81.1 (79.0) | 50.8 (53.6) | 33.7 (39.6) |
| rMAD SARAH-2 | 0.36 (0.34) | 0.30 (0.29) | 0.18 (0.20) | 0.11 (0.13) |
| RMSD HelioMont | 128.2 (113.4) | 101.2 (88.6) | 61.1 (51.7) | 26.9 (21.8) |
| MBD HelioMont | -7.0 (-7.3) | -6.7 (-6.8) | -7.0 (-7.5) | -6.3 (-7.7) |
| MAD HelioMont | 82.5 (72.0) | 66.7 (57.8) | 42.9 (35.1) | 22.2 (18.8) |
| rMAD HelioMont | 0.36 (0.33) | 0.29 (0.27) | 0.19 (0.17) | 0.07 (0.06) |

**Table 1.** Accuracy of the satellite-derived all-sky SSR from SARAH-2 and HelioMont averaged over all SMN ground stations (in brackets: averaged over all SACRaM and BSRN stations) for daytime periods in 2018 in W/m$^2$ for instantaneous SSR and for hourly, daily and monthly mean aggregation. For the definition of daytime periods including SZA aspects, see Section 4.

| **Day- and nighttime SSR** | **Instantaneous** | **Hourly mean** | **Daily mean** | **Monthly mean** |
|---|---|---|---|---|
| RMSD SARAH-2 | 87.2 (85.7) | 72.1 (71.0) | 27.8 (29.5) | 15.9 (14.4) |
| MBD SARAH-2 | -3.2 (-9.7) | -3.4 (-10.1) | 0.1 (-6.3) | -0.1 (-6.7) |
| MAD SARAH-2 | 40.2 (37.8) | 34.3 (32.0) | 20.6 (20.4) | 13.3 (12.7) |
| rMAD SARAH-2 | 0.45 (0.37) | 0.37 (0.29) | 0.17 (0.15) | 0.12 (0.12) |
| RMSD HelioMont | 81.2 (72.5) | 64.4 (56.3) | 25.6 (24.4) | 11.4 (8.8) |
| MBD HelioMont | -0.5 (-2.5) | -0.5 (-2.6) | -0.7 (-2.7) | -0.7 (-2.9) |
| MAD HelioMont | 35.8 (31.0) | 29.5 (25.1) | 17.7 (15.1) | 9.35 (7.6) |
| rMAD HelioMont | 0.45 (0.36) | 0.35 (0.28) | 0.16 (0.12) | 0.08 (0.07) |

**Table 2.** Accuracy of the satellite-derived all-sky SSR from SARAH-2 and HelioMont averaged over all SMN ground stations (in brackets: averaged over all SACRaM and BSRN stations) for day- and nighttime periods (24h periods) in 2018 in W/m$^2$.

To further quantify the uncertainty of the SSR measurements by the SMN pyranometers, we compare the SSR measured by the SMN and the BSRN pyranometers in Payerne. Both pyranometers are situated within 100 meters from each other at the MeteoSwiss meteorological station in Payerne. We find that the 10-minutes SSRs measured by the two pyranometers agree with each other within 2.38 W/m$^2$ of MAD, a Pearson correlation coefficient of 0.9995 and an MBD of 0.47 W/m$^2$ over all day- and nighttime periods of the year 2018. We perform the same analysis on the SACRaM station of Davos at which there is also another SMN pyranometer available, resulting in an MAD of 4.7 W/m$^2$ and an MBD of -2.0 W/m$^2$.
We repeat the analysis for only peak daytime SSR (10-13 UTC), finding that the SSR of the BSRN and SMN pyranometers at Payerne still match within less than 25 W/m$^2$ even for these peak periods. Specifically, the MAD in the instantaneous SSRs of the BSRN and SMN pyranometers is 6.8 W/m$^2$ and the MBD is 2.4 W/m$^2$, and the Pearson correlation is 0.998.
In summary, we consider our estimate of 25 W/m$^2$ a highly conservative upper limit for the uncertainty of the SSR measured by the SMN pyranometers, and find that the ground-measured SSR of the SMN constitutes a sufficiently accurate basis for investigating the accuracy and biases of the satellite-derived SSR. The uncertainty of the SACRaM and BSRN measurements are drastically smaller than the observed differences between satellite-estimated and ground-measured SSR, as listed in Tables 1 and 2. The SSR measurements of the SMN are more uncertain than the BSRN and SACRaM measurements due to possible soiling and non-compensation of thermal offset. However, this cannot explain the observed negative MBDs because the soiling and thermal offset effects would make the biases even larger in absolute value. Additional uncertainties in the ground-measured SSR are unlikely to explain the observed biases because the MDBs computed using the BSRN and SACRaM stations are similar to those using the SMN stations. The MBD of the HelioMont SSR in the SACRaM and BSRN stations is only moderately larger in absolute terms than for the SMN stations. By contrast, the MBD of the SARAH-2 SSR with regard to the SACRaM and BSRN stations is more than 10 W/m$^2$ larger than for the SMN stations (Table 1). This can be attributed to the fact that one SACRaM station (Davos) is located above 1000 m a.s.l. and is affected by SARAH-2's large underestimation of the SSR at higher altitudes. This effect is highlighted when the high-altitude stations (over 1000 m a.s.l.) are discarded as shown in Table 3, in which case the mean bias MBD of the instantaneous SSR of SARAH-2 decreases from -9.7 to -1.3 W/m$^2$ at the SACRaM and BSRN stations. The role of station altitude for SSR bias and accuracy is further detailed in Section 5.4.

Differences found when validating the satellite SSR estimates with ground observations come from a combination of satellite estimate uncertainty, ground-based measurement uncertainty and lack of correspondence between the spatial and temporal resolutions of the satellite and ground measurements. We have shown that the ground-based measurement uncertainty is small compared to the differences. The resolution-related uncertainties are discussed in sections 5.3 and 5.5.

| Day- and nighttime SSR<br>Low altitudes | Instantaneous | Hourly mean | Daily mean | Monthly mean |
|---|---|---|---|---|
| RMSD SARAH-2 | 69.9 (65.3) | 54.7 (50.9) | 19.7 (20.9) | 10.4 (6.1) |
| MBD SARAH-2 | 3.9 (-1.3) | 3.8 (-1.3) | 7.1 (2.2) | 7.0 (2.0) |
| MAD SARAH-2 | 31.3 (27.1) | 26.0 (22.1) | 14.9 (13.3) | 9.1 (5.1) |
| rMAD SARAH-2 | 0.41 (0.30) | 0.33 (0.24) | 0.13 (0.10) | 0.09 (0.06) |
| RMSD HelioMont | 71.2 (61.5) | 54.6 (46.1) | 21.1 (20.6) | 8.3 (5.5) |
| MBD HelioMont | -1.9 (-3.2) | -1.9 (-3.1) | -2.2 (-3.4) | -2.1 (-3.2) |
| MAD HelioMont | 31.1 (25.7) | 25.1 (20.3) | 14.4 (12.1) | 6.7 (4.6) |
| rMAD HelioMont | 0.41 (0.32) | 0.32 (0.25) | 0.14 (0.11) | 0.07 (0.05) |

**Table 3.** Accuracy of the satellite-derived all-sky SSR from SARAH-2 and HelioMont averaged over all SMN ground stations (in brackets: averaged over the SACRaM and BSRN stations of Locarno and Payerne) at altitudes below 1000 m a.s.l. for day- and nighttime periods (24h periods) in 2018 in W/m$^2$.

## 5.3 Role of time resolution

We further investigate the impact of the time resolution on the bias estimates of the instantaneous satellite SSR. To this end, we also exploit 1-minute average SSR ground measurements from the three SACRaM and BSRN stations. SSR measurements are available at 1-minute resolution in addition to the usual 10-minute mean SSR. We find that the inaccuracy of the satellite SSR increases with regard to the ground-measured SSR when going from 10- to 1-minute aggregation. For instance, the instantaneous MADs of the HelioMont SSR are 41.6, 25.2, 26.2 W/m$^2$ for Davos, Locarno and Payerne, respectively, in the 10-minute SSR, and they increase to 44.9, 27.0 and 28.3 W/m$^2$ when comparing HelioMont to the 1-minute SSR at the three stations.

The largest inaccuracies and biases are observed around noontime which is when PV production typically peaks. During these peak periods, the satellite-derived instantaneous SSR deviates from the ground-measured SSR with an MAD of 110.4 and 99.6 W/m$^2$ for SARAH-2 and HelioMont, respectively, as shown in Table A2 (supplementary material).

## 5.4 Role of site altitude for accuracy

Figure 1 illustrates the RMSDs for all the SMN ground stations for different time aggregations. Each dot represents one station with the dot size proportional to the station altitude and the color indicating the RMSD. As can be seen, the uncertainty (RMSDs) in the satellite SSR estimates is larger at higher altitudes and in mountainous areas. This altitude effect is somewhat more severe in the case of SARAH-2. Even in the monthly mean, the RMSD in the satellite-derived SSR exceeds 100 W/m$^2$ at multiple Alpine stations in the case of SARAH-2.

Figure 2 shows that an altitude dependence not only exists for SSR accuracy but also for SSR biases. The satellite estimates of the instantaneous SSR of SARAH-2 and HelioMont indicate significant under- and overestimation across many of the 133 ground stations. The biases are particularly pronounced in the mountainous parts of Switzerland. SARAH-2 significantly underestimates the SSR at altitudes above approximately 1000 m, as demonstrated in the Swiss Alps. The HelioMont SSR biases are less pronounced than the ones of SARAH-2 and present a more station-specific pattern, with a tendency to underestimate the SSR in the northern parts of the Alps and to overestimate it in the southern parts of the Alps. Figure A1 (supplementary material) presents the distributions of the site-specific biases, also showing a strong underestimation of the SSR by SARAH-2 at high altitudes and more moderate tendencies of SARAH-2 to overestimate the SSR at low altitudes and of HelioMont to underestimate the SSR at low altitudes.

Figure 3 illustrates the seasonal and altitude dependence of the biases in the satellite SSR estimates. As seen previously, there is a tendency towards negative biases at higher elevations in the case of SARAH-2. At lower elevations, it confirms that the biases tend to be positive in SARAH-2 and negative in HelioMont. In the summer season, the spatial bias pattern appears to be more similar among SARAH-2 and HelioMont than in the other seasons. The biases in the daytime SSR of SARAH-2 exhibit a seasonal cycle which is most pronounced at higher altitudes and in the mountainous areas. The mean bias (MBD) indicates large negative values there, which significantly exceed 100 W/m$^2$ in winter and spring. One possible explanation of the observed seasonal cycle in the biases is a misinterpretation of snow cover as clouds by the SARAH-2 algorithm. An example of this SSR underestimation behavior is presented in Figures A2 and A3, illustrating how SARAH-2 misses out on the SSR for part of the Alpine region, therefore resulting in a strong negative bias. The identified SSR underestimation can, in principle, also have other causes, e.g., deficiencies in the clear-sky model inputs, such as in the assumed trace gas concentrations at high altitudes. Identifying the actual causes of SSR biases in the SARAH-2 and HelioMont products is beyond the scope of this study.

Figure 4 shows the mean biases of the SSR over the course of the year 2018. The stations are arranged by altitudes. They are straddling an altitude range of more than 3.3 km from the Lago Maggiore (Magadino: 203 m a.s.l., 46.16°N, 8.93°E) to the Bernese Alps (Jungfraujoch: 3571 m a.s.l., 46.55°N, 7.99°E). Gray pixels indicate missing data due to the site-specific SZA filtering applied. The SSR biases discussed above are again clearly apparent in this representation of the data. Moreover, Figure 4 illustrates the systematic underestimation of the SARAH-2 SSR in the first four and the last two months of 2018. The most abrupt change in the SSR mean bias in 2018 occurs simultaneously along a range of high altitudes at the end of October 2018, as shown in the upper right panel of Figure 4. This sudden increase in the MBD of the SARAH-2 SSR coincides with the first snow fall on 27 October. A possible plausible explanation of the sudden strong increase in underestimation bias at high altitudes at the end of October, shown in Figure 4, is that it was caused by a misclassification of snow cover as clouds. Snow cover related biases of SARAH-2 have also been mentioned in previous studies, e.g. Buffat et al., 2015; Pfeifroth et al., 2018; Babar et al., 2019.

Moreover, the SARAH-2 and HelioMont instantaneous SSRs are characterized by major inaccuracies in summer time, as shown by the RMSD, but the associated biases are less systematic in terms of their signs. The large SSR uncertainties in summertime are at least in part explained by the fact that the SSR tends to be larger in summer.

Systematic mismatches between ground-measured SSR and satellite-estimated SSR can be due to a number of reasons. In terms of the pyranometer measurement process, error sources include calibration uncertainties, electronics-related uncertainties as well as leveling- and soiling-induced biases (Vuilleumier et al., 2014). Yet, as discussed above, the total amount of measurement uncertainty of ground-based SSR is at most 18 W/m$^2$ in the case of the BSRN and SACRaM stations, and no more than 25 W/m² for the pyranometers of the SMN stations. Therefore, the magnitude of those uncertainties cannot explain the observed differences between the ground measurements and the satellite estimates.

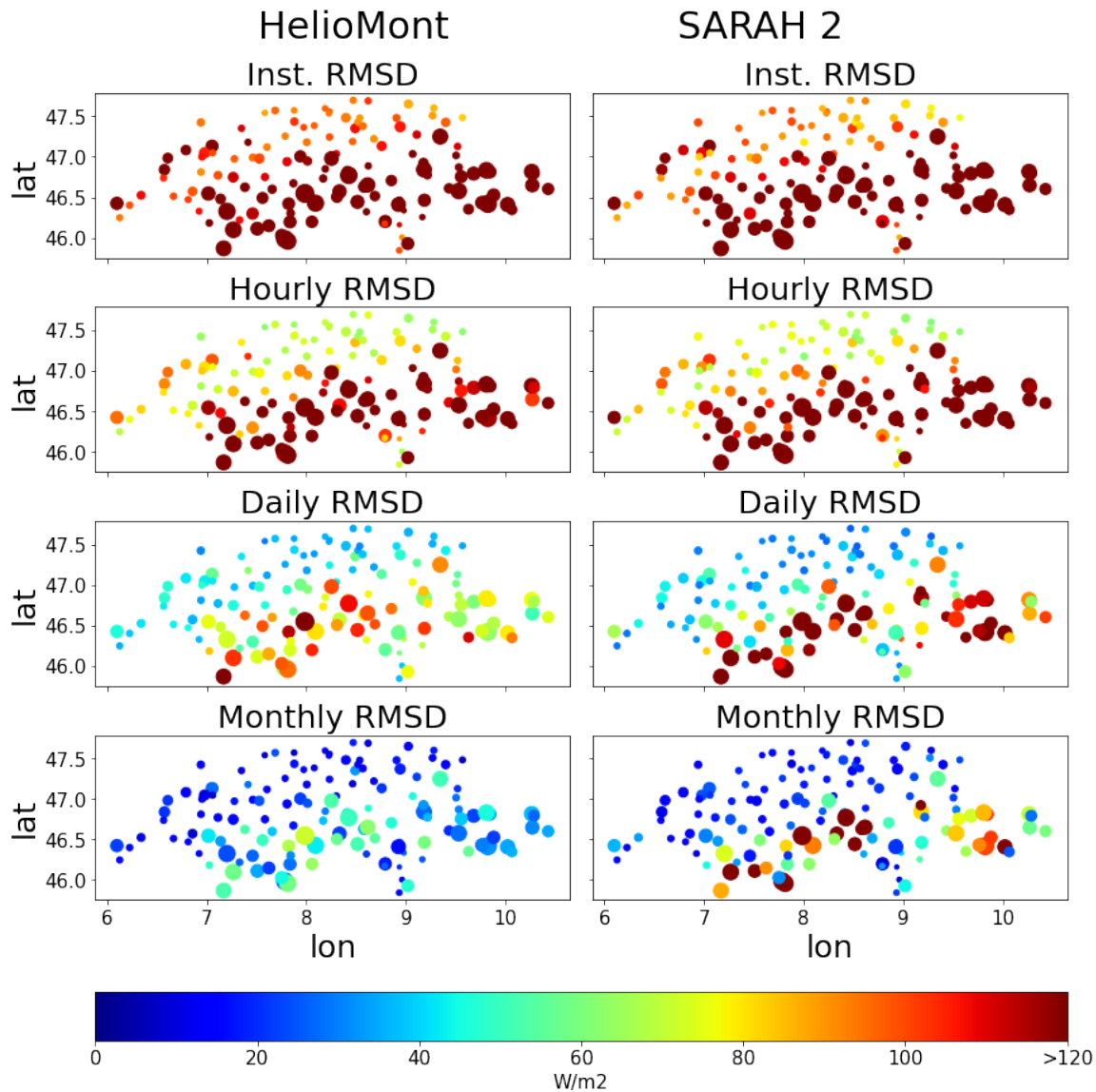

**Figure 1.** RMSD between the satellite-derived and the ground-measured all-sky SSR in Switzerland for daytime periods at the SMN stations in 2018. The RMSDs are provided for instantaneous measurements and for hourly, daily and monthly mean aggregation. Each dot indicates the location of a SMN ground station and the associated RMSD. The dot sizes are proportional to the station altitudes.

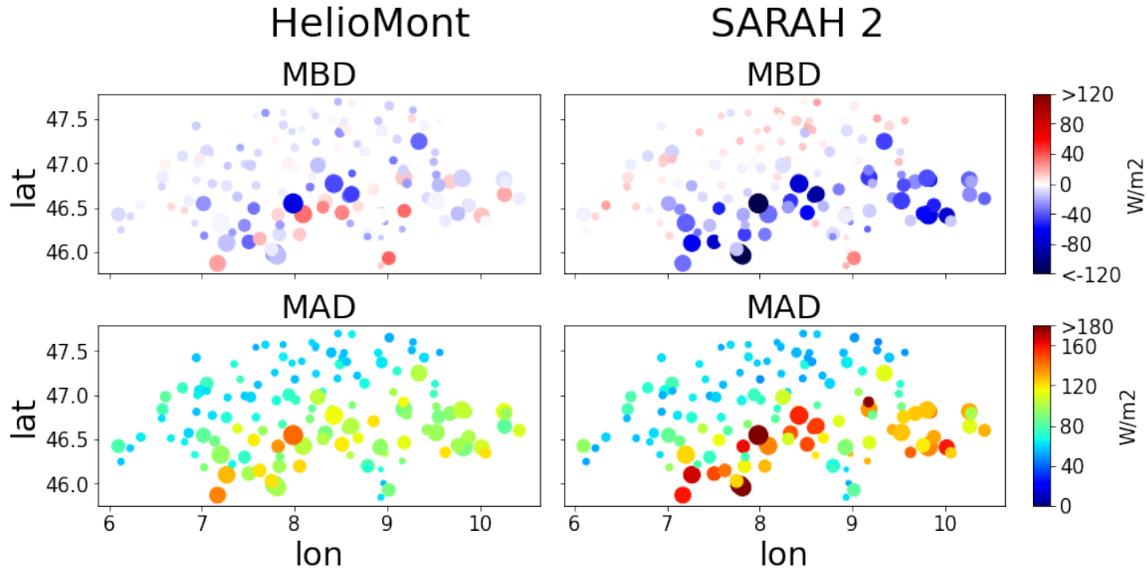

**Figure 2.** Mean biases and MADs of the satellite estimates of the instantaneous daytime all-sky SSR at all SMN ground stations in 2018. For each station, the biases are averaged over all instantaneous time steps in 2018. Negative MBDs indicate that the satellite SSR underestimates the ground measurements. Each dot indicates a pyranometer of a SMN ground station. The dot sizes are proportional to the station altitudes. The metrics are computed using only timestamps having solar zenith angles lower than the site-specific threshold $SZA^{max}$.

## 5.5 Role of spatial resolution

We also investigate the potential error made by comparing SSR point measurements to satellite pixels covering a 0.02˚x0.02˚ (HelioMont) or 0.05˚x0.05˚ (SARAH-2) surface area. The direct comparison of an SSR point measurement with a spatially extended satellite SSR estimate pixel may induce a spatial representativeness error (Li et al., 2005; Hakuba et al., 2013; Li, 2014; Schwarz et al., 2018). This is because the satellite pixel does not capture the sub-pixel variability, whereas the point measurement does not cover the full ground area of the satellite-derived SSR pixel, so the ground measurement has only limited spatial representativeness.

In our study, 133 pixels in our satellite SSR maps have a corresponding ground station, whereas most of our satellite pixels have no ground station in their footprint area. Only two pixels comprise two ground stations each, namely the pixel associated to the SMN and BSRN stations at Payerne and the pixel associated to the SMN and SACRaM stations at Davos. However, no pixel comprises more than two ground stations. A much larger number of ground stations in each pixel would be needed to investigate the spatial representativeness of the ground stations. This requirement for quantifying the spatial representativeness cannot be realistically satisfied for the SwissMetNet, just as for any other operational meteorological network, due to the high resolution of the satellite images used in this study. Therefore, we make use of the spatial representativeness errors derived in previous work.

In particular, Huang et al., 2016 studied the representativeness errors for ground measurements from 16 pyranometers distributed across a kilometer-size pixel area in Northwest China. The spatial representativeness error was computed as the difference between the single station measurement and the average value of the instantaneous SSR measurements among all the stations inside the area considered. They found representativeness errors (RMSDs) equal to 21.9 W/m² and 40.2 W/m² for 1x1 km² and 5x5 km² areas, respectively. They concluded that the error derived from an inadequate representativeness of point scale measurements increases the RMSD by 13.4% for instantaneous satellite SSR products with 0.05˚x0.05˚ spatial resolution. They quantified this value by using the average measurements among all pyranometers instead of single-site values.

Applying the results of Huang et al., 2016 to the present study, the RMSD of the 0.05°x0.05° instantaneous satellite SSR from SARAH-2 needs to be corrected by about 13.4% which results in a reduction by 11.8% of the RMSDs we report in Figure 1. When correcting for the representativeness error in the HelioMont instantaneous SSR, the decrease would be less than 11.8% because the HelioMont SSR comes at a higher spatial resolution.

Differences between satellite and ground SSR originate from a combination of satellite SSR uncertainty, ground-based measurement uncertainty, and uncertainty related to differences in temporal and spatial resolution between the point ground-based measurements and satellite SSR pixel estimates. Our results show that a significant inaccuracy in the satellite-derived SSR remains even after accounting for the potential uncertainties of the ground measurements and the temporal and spatial resolution differences.

### 5.6 Consistency with earlier results

The reported differences between satellite-based and in situ SSR may look surprisingly large at first glance. As explained above, this is essentially due to the data selection, tailored with PV applications in mind: the focus on short non-aggregated (instantaneous) time scales instead of, e.g., monthly means; the focus on day-time and high solar elevation (SZA) as peak PV production time; and the focus on high-altitude sites, given their potential for wintertime PV production in mountain terrain. Our results are consistent with previous studies when these selection criteria are omitted in favor of a conventional bias and accuracy evaluation. Müller et al., 2015 analyzed the SARAH-2 accuracy based on 15 BSRN stations in mainly European countries, including day- and nighttime values between 1992-2013, and arrived at inaccuracies (MAD) in the daily and monthly-mean SSR estimates of 12 and 5.5 W/m$^2$. This is in line with the 11.5 and 5.5 W/m$^2$ that we obtain at the BSRN station in Payerne. The median hourly RMSD value for the SSR of SARAH is comparable to the one by Greuell et al., 2013 for European BSRN stations, 61.4 vs. 65 W/m$^2$. Furthermore, Riihelä et al., 2015, found the SARAH monthly-mean and daily-mean RMSD across all stations of the Swedish and Finnish meteorological networks to be 8.3 and 17.0 W/m$^2$, while for the SMN network the values are 15.9 and 27.8 W/m$^2$; but considering only low altitude SMN stations (<1000 m a.s.l.), the corresponding uncertainties are 10.4 and 19.7 W/m$^2$, as shown in Table 3. Analyzing the MAD of the monthly-mean SSR of SARAH-2, Babar et al., 2019 found an MAD of 5.0 W/m$^2$ across 31 Scandinavian stations, which is comparable to the MAD for our low altitude SACRaM and BSRN stations of 5.1 W/m$^2$ and consistent with the MAD of 9.1 W/m$^2$ across the SMN stations. Lastly, Castelli et al., 2014 reported mean absolute deviations in the HelioMont hourly-mean SSR of 40 and 52 W/m$^2$ at the ground stations of Payerne and Davos, Switzerland, respectively. In Payerne and Davos, our study finds similar daytime hourly mean absolute deviations (41.8 and 66.6 W/m$^2$). When studying systematic under- and overestimations, Castelli et al., 2014 reported mean biases of 2 and -6 W/m$^2$ in hourly-mean SSR at these stations, which is in line with our findings. None of the above studies investigated the accuracy and biases of instantaneous intra-hour SSR estimates derived from Meteosat.

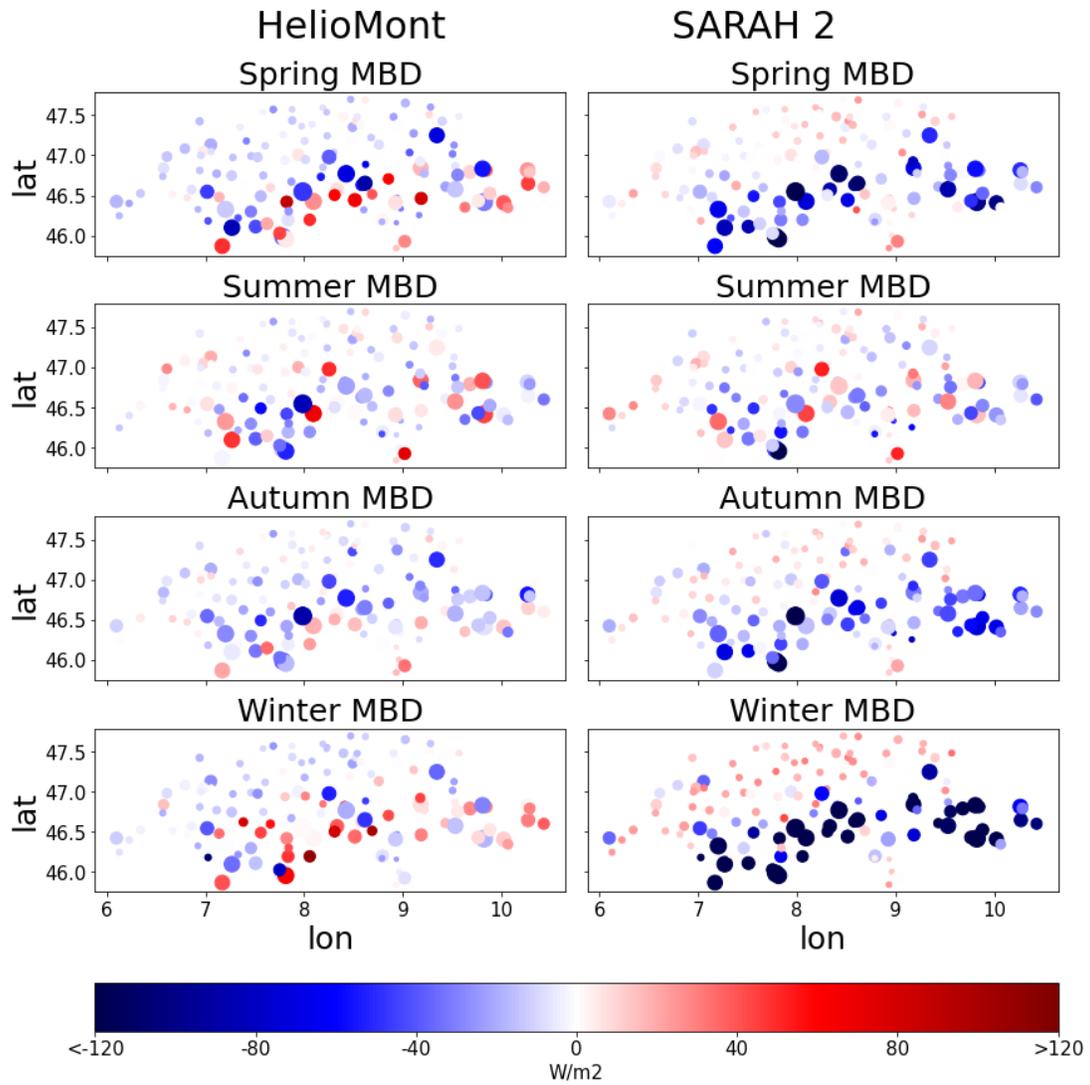

**Figure 3.** Mean biases of the satellite-derived daytime all-sky SSR by season. Each dot indicates a pyranometer of a SMN ground station. The dot sizes are proportional to the station altitudes. Spring comprises March, April, May. Summer includes June, July, August, and so forth.

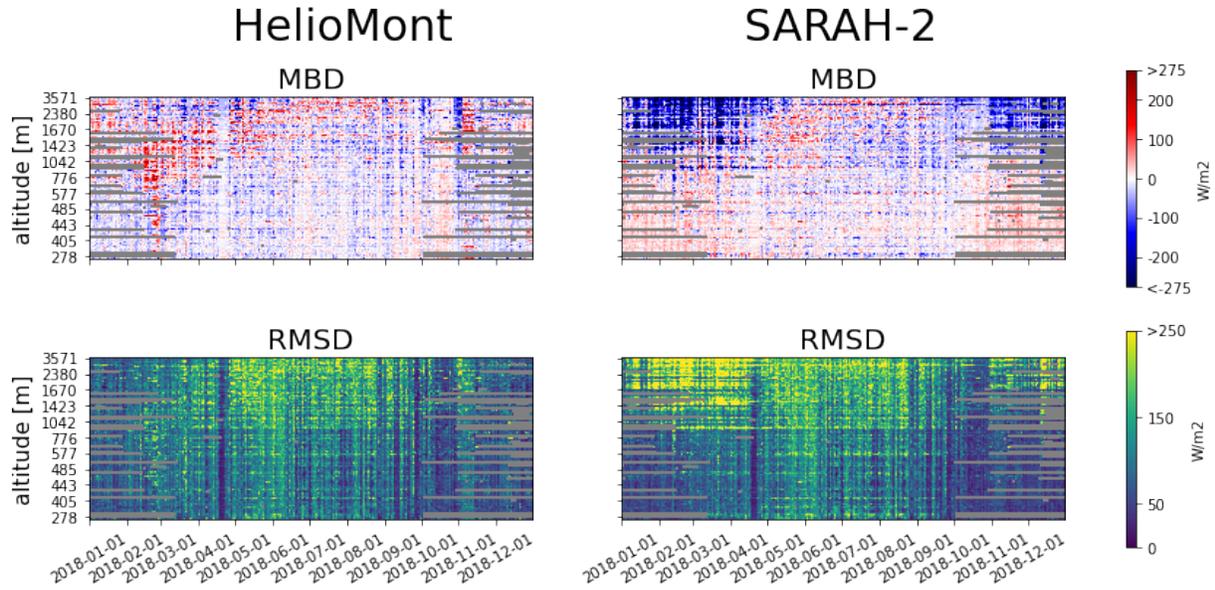

**Figure 4.** Mean biases and RMSDs of the satellite-derived estimates of the daytime all-sky SSR for all days of 2018. The SMN ground stations are sorted along the vertical axis according to their altitude. Every row in each subpanel corresponds to a SMN station, every column corresponds to a day. Each pixel indicates the mean bias (or RMSD) of the instantaneous satellite SSR at that ground station on that day. Gray pixels are missing data due to the applied SZA filter to avoid time periods in which the sun was behind the horizon.

## 6. Conclusions

Accurate estimates of the intra-hour and intra-day surface solar radiation are crucial for applications such as real-time estimation and forecasting of the photovoltaic power generation. SSR estimates are of great importance in the operational activities of grid operators and power traders to optimize power generation revenues and maintain power grid stability. We investigated the accuracy of intra-hour and intra-day SSR estimates derived from Meteosat SEVIRI with the SARAH-2 and HelioMont algorithms. SSR uncertainties and biases were studied based on 136 ground stations in Switzerland at altitudes between 200 to 3570 m a.s.l. in 2018.

Our study confirms the low uncertainties of 3-6 W/m$^2$ in the monthly-mean Heliosat SSR (Müller et al., 2015) for the BSRN station Payerne. We demonstrated that the SSR uncertainties grow by one to two orders of magnitude when going to intra-hour and intra-day time scales and focusing on daytime periods. We found significant inaccuracies in the SARAH-2 and HelioMont SSR products, with daytime mean, hourly mean and instantaneous SSR RMSDs of 70.2, 116.7, 136.9 W/m$^2$ in SARAH-2, and 61.1, 101.2 and 128.2 W/m$^2$ in the HelioMont SSR. Moreover, we found that the SARAH-2 SSR significantly underestimates the solar resources at higher altitudes and mountain regions in the winter half-year. The biases in the instantaneous SSR amount to more than one hundred W/m$^2$ at multiple stations. As a possible explanation, SARAH-2 may be systematically misinterpreting snow cover as clouds, which could explain the drastic underestimation of the SSR in regions affected by seasonal snow cover. We also found a more moderate overestimation of the SARAH-2 SSR at lower altitudes and a moderate underestimation bias of the HelioMont SSR at lower altitudes.

The inaccuracies and biases in the satellite-derived SSR are largest during peak daytime periods (Table A2) in both satellite products, as expected. The accuracy of both products increases when including also low altitude stations, non-peak daytime periods as well as nighttime periods (Tables 1-3 and A2). Temporal aggregation also improves the satellite SSR accuracies.

Our study focuses on instantaneous SSR estimates during daytime periods, applying an SZA filtering to remove shadowing effects that favors noontime periods. In addition, and in contrast to previous studies, we also include medium- and high-altitude stations in our analysis. Each of these four conditions (no time aggregation, daytime only, favoring peak daytime periods, including high altitudes) tends to increase the biases and uncertainties of the satellite-derived SSRs. This explains the remarkably large inaccuracies in the satellite-derived daytime SSR (Tables 1 and A2) which exceed the inaccuracies reported in previous studies. Previous studies focused on time-aggregated SSR at lower altitudes and included nighttime periods. The uncertainties in the instantaneous daytime SSR estimates of SARAH-2 and HelioMont exceed the pyranometer and resolution-related uncertainties. Our results demonstrate that the instantaneous SSR estimates of SARAH-2 and HelioMont are only imperfect proxies of in situ measured SSR in that they are significantly less accurate and exhibit biases across the SMN stations.

Our findings shed new light on the solar resources in elevated regions which are frequently affected by winter-time snow cover, such as the Alps. Photovoltaic resources could indeed become one of the most relevant renewable energy sources in these regions alongside hydropower. On the one hand, more solar resources are available at these higher altitudes due to the shorter atmospheric path of the solar beam, the associated lower aerosol burden and water vapor column, and correspondingly reduced beam attenuation. In addition, elevated regions are usually also less affected by cloud cover in Switzerland in wintertime due to their position above the low-level stratiform clouds often persisting in winter. On the other hand, Alpine hydropower dams and reservoirs offer large unused surfaces suitable for PV system operation. PV power is particularly valuable in wintertime when overcast sky conditions dominate the lowlands while higher regions remain relatively unaffected by clouds (Kahl et al., 2019; Dujardin et al., 2022).

We confirmed the accuracy of the Heliosat SSR estimates at monthly-mean time scales in both SSR products. This time resolution and the associated SSR accuracy can be sufficient for studies focused on long-term climatological trends in surface solar radiation, such as studies of global dimming and brightening (e.g., Wild, 2009). However, at intra-hour and intra-day observation periods, the satellite-derived SSR estimates are subject to major uncertainties and biases. As these time scales are relevant for applications such as short-term forecasting of solar resources and PV power generation, further research should investigate how the existing biases can be reduced and how more accurate SSR estimates can help improve the quality of the short-term applications of satellite-derived SSR. This includes, in particular, approaches to deal with the spatial and temporal heterogeneity of SSR bias and accuracy patterns, notably correcting SSR underestimation biases at higher altitudes, as illustrated and quantified in this study. This might be achieved by including accurate additional information on cloud cover or snow cover in the bias correction approach. Our study highlights the need for further research into bias correction methods for the satellite-derived SSR. We expect that our results can trigger and facilitate the development of methods for more accurate and unbiased intra-hour and intra-day solar resource estimates.

**Acknowledgements**
The authors thank Christian Félix and Yves-Alain Roulet from the Federal Office of Meteorology and Climatology MeteoSwiss for providing the SSR pyranometer measurements of the SwissMetNet stations. They also thank Luca Modolo from MeteoSwiss for providing site-specific profiles of the horizon seen by each SMN station. The authors also wish to express their gratitude to Reto Stöckli, Anke Tetzlaff and Quentin Bourgeois from MeteoSwiss and to Jörg Trentmann from the German Weather Service DWD for providing the HelioMont and SARAH-2 SSR estimates and for fruitful discussions. A.C. and A.M. gratefully acknowledge funding by the Swiss National Science Foundation under grant number 200021_200654. The SMN data used in our study are available through https://gate.meteoswiss.ch/idaweb. The SARAH-2 data are available via https://wui.cmsaf.eu/safira/. The HelioMont data are licensed and can be obtained from the MeteoSwiss customer service via https://www.meteoswiss.admin.ch/home/form/customer-service.html. Our code is available on https://github.com/albertocarpentieri/BiasAssessment.

**Supplementary Material**

| Station | Abbreviation | Altitude m a.s.l. | Longitude °N | Latitude °E |
|---|---|---|---|---|
| Magadino/Cadenazzo | MAG | 203 | 46.16 | 8.93 |
| Lugano | LUG | 273 | 46 | 8.96 |
| Biasca | BIA | 278 | 46.34 | 8.98 |
| Basel/Binningen | BAS | 316 | 47.54 | 7.58 |
| Grono | GRO | 324 | 46.26 | 9.16 |
| Beznau | BEZ | 326 | 47.56 | 8.23 |
| Würenlingen/PSI | PSI | 334 | 47.54 | 8.23 |
| Leibstadt | LEI | 341 | 47.6 | 8.19 |
| Möhlin | MOE | 343 | 47.57 | 7.88 |
| Stabio | SBO | 351 | 45.84 | 8.93 |
| Locarno-Monti[+] | OTL | 367 | 46.17 | 8.79 |
| Gösgen | GOE | 380 | 47.36 | 7.97 |
| Aigle | AIG | 381 | 46.33 | 6.92 |
| Buchs/Aarau | BUS | 387 | 47.38 | 8.08 |
| Altenrhein | ARH | 398 | 47.48 | 9.57 |
| Vevey/Corseaux | VEV | 405 | 46.47 | 6.82 |
| Genève/Cointrin | GVE | 411 | 46.25 | 6.13 |
| Cevio | CEV | 417 | 46.32 | 8.6 |
| Hallau | HLL | 419 | 47.7 | 8.47 |
| Wynau | WYN | 422 | 47.26 | 7.79 |
| Zürich/Kloten | KLO | 426 | 47.48 | 8.54 |
| Grenchen | GRE | 428 | 47.18 | 7.42 |
| Cressier | CRM | 430 | 47.05 | 7.06 |
| Mathod | MAH | 435 | 46.74 | 6.57 |
| Altdorf | ALT | 438 | 46.89 | 8.62 |
| Schaffhausen | SHA | 438 | 47.69 | 8.62 |
| Delémont | DEM | 439 | 47.35 | 7.35 |
| Güttingen | GUT | 440 | 47.6 | 9.28 |
| Cham | CHZ | 443 | 47.19 | 8.46 |
| Zürich/Affoltern | REH | 444 | 47.43 | 8.52 |
| Mosen | MOA | 453 | 47.24 | 8.23 |
| Luzern | LUZ | 454 | 47.04 | 8.3 |
| Pully | PUY | 456 | 46.51 | 6.67 |
| Vaduz | VAD | 457 | 47.13 | 9.52 |
| Nyon/Changins | CGI | 458 | 46.4 | 6.23 |
| Giswil | GIH | 471 | 46.85 | 8.19 |
| Mühleberg | MUB | 480 | 46.97 | 7.28 |
| Evionnaz | EVI | 482 | 46.18 | 7.03 |
| Sion | SIO | 482 | 46.22 | 7.33 |
| Koppigen | KOP | 485 | 47.12 | 7.61 |
| Wädenswil | WAE | 485 | 47.22 | 8.68 |
| Neuchâtel | NEU | 485 | 47 | 6.95 |
| Payerne | PAY | 490 | 46.81 | 6.94 |
| St. Chrischona | STC | 493 | 47.57 | 7.69 |
| Bad Ragaz | RAG | 497 | 47.02 | 9.5 |
| Bischofszell/Sitterdorf | BIZ | 507 | 47.51 | 9.27 |
| Glarus | GLA | 517 | 47.03 | 9.07 |
| Egolzwil | EGO | 522 | 47.18 | 8 |
| Aadorf/Tänikon | TAE | 539 | 47.48 | 8.9 |
| Bern/Zollikofen | BER | 553 | 46.99 | 7.46 |
| Zürich/Fluntern | SMA | 556 | 47.38 | 8.57 |
| Chur | CHU | 556 | 46.87 | 9.53 |
| Thun | THU | 570 | 46.75 | 7.59 |
| Acquarossa | COM | 575 | 46.46 | 8.94 |
| Interlaken | INT | 577 | 46.67 | 7.87 |
| Meiringen | MER | 589 | 46.73 | 8.17 |
| Fahy | FAH | 596 | 47.42 | 6.94 |
| Rünenberg | RUE | 611 | 47.43 | 7.88 |
| Ebnat-Kappel | EBK | 623 | 47.27 | 9.11 |
| Visp | VIS | 639 | 46.3 | 7.84 |
| Fribourg | GRA | 651 | 46.77 | 7.11 |
| Bière | BIE | 684 | 46.52 | 6.34 |
| Ilanz | ILZ | 698 | 46.78 | 9.22 |
| Salen-Reutenen | HAI | 719 | 47.65 | 9.02 |
| Schüpfheim | SPF | 744 | 46.95 | 8.01 |
| Langnau i.E. | LAG | 744 | 46.94 | 7.81 |
| Frutigen | FRU | 756 | 46.6 | 7.66 |
| St. Gallen | STG | 776 | 47.43 | 9.4 |
| Boltigen | BOL | 820 | 46.62 | 7.38 |
| Oron | ORO | 828 | 46.57 | 6.86 |
| Lägern | LAE | 845 | 47.48 | 8.4 |
| Uetliberg | UEB | 854 | 47.35 | 8.49 |
| Einsiedeln | EIN | 911 | 47.13 | 8.76 |
| Bantiger | BAN | 942 | 46.98 | 7.53 |
| Elm | ELM | 958 | 46.92 | 9.18 |
| Andeer | AND | 987 | 46.61 | 9.43 |
| Piotta | PIO | 990 | 46.51 | 8.69 |
| La Chaux-de-Fonds | CDF | 1017 | 47.08 | 6.79 |
| Château-d'Oex | CHD | 1028 | 46.48 | 7.14 |
| Engelberg | ENG | 1036 | 46.82 | 8.41 |
| Plaffeien | PLF | 1042 | 46.75 | 7.27 |
| La Brévine | BRL | 1050 | 46.98 | 6.61 |
| Robbia | ROB | 1078 | 46.35 | 10.06 |
| Vicosoprano | VIO | 1089 | 46.35 | 9.63 |
| Hörnli | HOE | 1133 | 47.37 | 8.94 |
| Chaumont | CHM | 1136 | 47.05 | 6.98 |
| Disentis | DIS | 1197 | 46.71 | 8.85 |
| Bullet/La Frétaz | FRE | 1205 | 46.84 | 6.58 |
| Scuol | SCU | 1304 | 46.79 | 10.28 |
| Adelboden | ABO | 1321 | 46.49 | 7.56 |
| Ulrichen | ULR | 1346 | 46.5 | 8.31 |
| Val Müstair | SMM | 1386 | 46.6 | 10.43 |
| Napf | NAP | 1404 | 47 | 7.94 |
| Montana | MVE | 1423 | 46.3 | 7.46 |
| Andermatt | ANT | 1435 | 46.63 | 8.58 |
| Simplon-Dorf | SIM | 1465 | 46.2 | 8.06 |
| Blatten | BLA | 1538 | 46.42 | 7.82 |
| Valbella | VAB | 1568 | 46.76 | 9.55 |
| Mottec | MTE | 1580 | 46.15 | 7.62 |
| Davos[+] | DAV | 1594 | 46.81 | 9.84 |
| Chasseral | CHA | 1594 | 47.13 | 7.05 |
| Mt. Generoso | GEN | 1600 | 45.93 | 9.02 |
| Grächen | GRC | 1605 | 46.2 | 7.84 |
| Zermatt | ZER | 1638 | 46.03 | 7.75 |
| S. Bernardino | SBE | 1639 | 46.46 | 9.18 |
| Cimetta | CIM | 1661 | 46.2 | 8.79 |
| La Dôle | DOL | 1670 | 46.42 | 6.1 |
| Samedan | SAM | 1709 | 46.53 | 9.88 |
| Segl-Maria | SIA | 1804 | 46.43 | 9.76 |
| Evolène | EVO | 1825 | 46.11 | 7.51 |
| Arosa | ARO | 1878 | 46.79 | 9.68 |
| Robièi | ROE | 1898 | 46.44 | 8.51 |
| Buffalora | BUF | 1971 | 46.65 | 10.27 |
| Le Moléson | MLS | 1974 | 46.55 | 7.02 |
| Grimsel | GRH | 1980 | 46.57 | 8.33 |
| Pilatus | PIL | 2105 | 46.98 | 8.25 |
| Matro | MTR | 2171 | 46.41 | 8.92 |
| Berninapass | BEH | 2260 | 46.41 | 10.02 |
| Gütsch | GUE | 2286 | 46.65 | 8.62 |
| Naluns | NAS | 2380 | 46.82 | 10.26 |
| Crap Masegn | CMA | 2468 | 46.84 | 9.18 |
| St-Bernard | GSB | 2472 | 45.87 | 7.17 |
| Säntis | SAE | 2501 | 47.25 | 9.34 |
| Piz Martegnas | PMA | 2668 | 46.58 | 9.53 |
| Weissfluhjoch | WFJ | 2691 | 46.83 | 9.81 |
| Les Attelas | ATT | 2734 | 46.1 | 7.27 |
| Monte Rosa | MRP | 2885 | 45.96 | 7.81 |
| Eggishorn | EGH | 2892 | 46.43 | 8.09 |
| Les Diablerets | DIA | 2964 | 46.33 | 7.2 |
| Titlis | TIT | 3040 | 46.77 | 8.43 |
| Gornergrat | GOR | 3129 | 45.98 | 7.79 |
| Piz Corvatsch | COV | 3294 | 46.42 | 9.82 |
| Jungfraujoch | JUN | 3571 | 46.55 | 7.99 |
| BSRN Payerne*,+ | - | 491 | 46.82 | 6.94 |

**Table A1.** Ground stations of the SwissMetNet, the SACRaM (+) and the BSRN (*) that provided pyranometer measurements analyzed in this work.

| **Peak daytime SSR biases (10 - 13 UTC)** | **Instantaneous** | **Hourly mean** | **Peak daytime mean** | **Monthly peak daytime mean** |
|---|---|---|---|---|
| RMSD SARAH-2 | 158.8 (159.0) | 130.4 (133.1) | 110.3 (114.5) | 59.0 (66.3) |
| MBD SARAH-2 | -32.5 (-48.3) | -32.2 (-48.1) | -32.2 (-48.1) | -32.6 (-48.1) |
| MAD SARAH-2 | 110.4 (109.2) | 92.4 (93.5) | 79.1 (81.6) | 48.8 (56.4) |
| rMAD SARAH-2 | 0.35 (0.31) | 0.28 (0.26) | 0.23 (0.22) | 0.12 (0.13) |
| RMSD HelioMont | 149.9 (133.4) | 119.5 (105.3) | 99.3 (86.3) | 43.9 (37.7) |
| MBD HelioMont | -14.1 (-17.1) | -14.8 (-17.9) | -14.8 (-17.9) | -14.9 (-18.1) |
| MAD HelioMont | 99.6 (84.8) | 81.6 (68.2) | 69.2 (56.4) | 36.6 (30.0) |
| rMAD HelioMont | 0.38 (0.30) | 0.31 (0.24) | 0.26 (0.20) | 0.10 (0.07) |

**Table A2.** Accuracy of the satellite-derived all-sky SSR from SARAH-2 and HelioMont averaged over all SMN ground stations (in brackets: averaged over all SACRaM and BSRN stations) for peak daytime periods (between 10-13 UTC) in 2018 in W/m$^2$ for instantaneous SSR and for hourly, daily and monthly mean aggregation. The SSR peaks between 11.15 and 11.45 UTC at all sites and all times in 2018, so 11.30 UTC +/- 90 min is chosen to define the peak clear-sky SSR time of day.

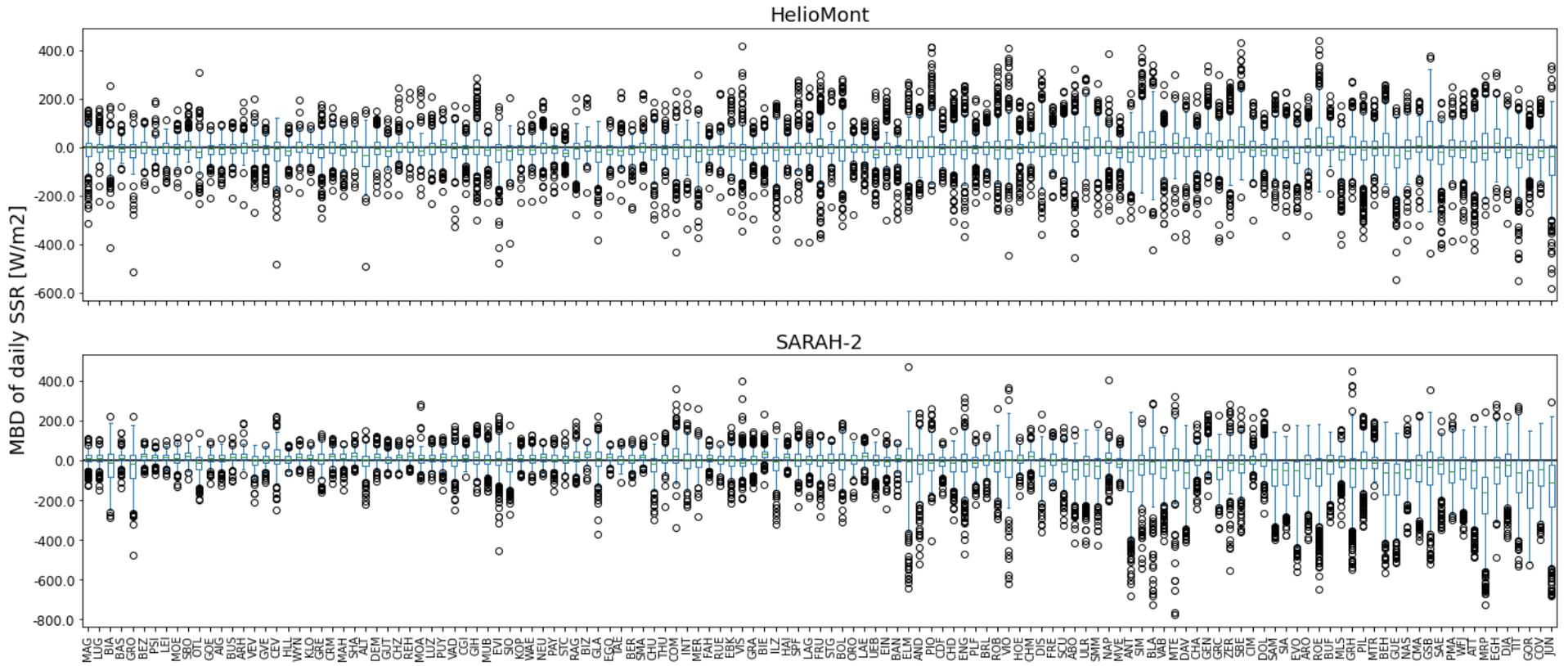

**Figure A1.** Boxplots of the mean biases (MBD) of the daily-mean all-sky daytime SSR from HelioMont and SARAH-2 with regard to each SMN ground station in 2018. The stations are sorted along the horizontal axis by their altitudes, with the lowest one (MAG) on the left-hand side and the highest one (JUN) son the right-hand side. Table A1 provides the names, locations and altitudes of the stations for the station name abbreviations on the horizontal axis of this plot. The spread of MBD distribution increases with altitude. For SARAH-2, the MBD exhibits a negative trend with increasing altitude, exhibiting a strong underestimation of the daily-mean SSR estimates at higher sites towards the right-hand side of the plot. We present the distribution of the MBD in the daily-mean SSR, rather than in the instantaneous or hourly-mean SSR, because the boxplots of the latter distributions are too noisy to visualize and interpret due to their large amounts of outliers.

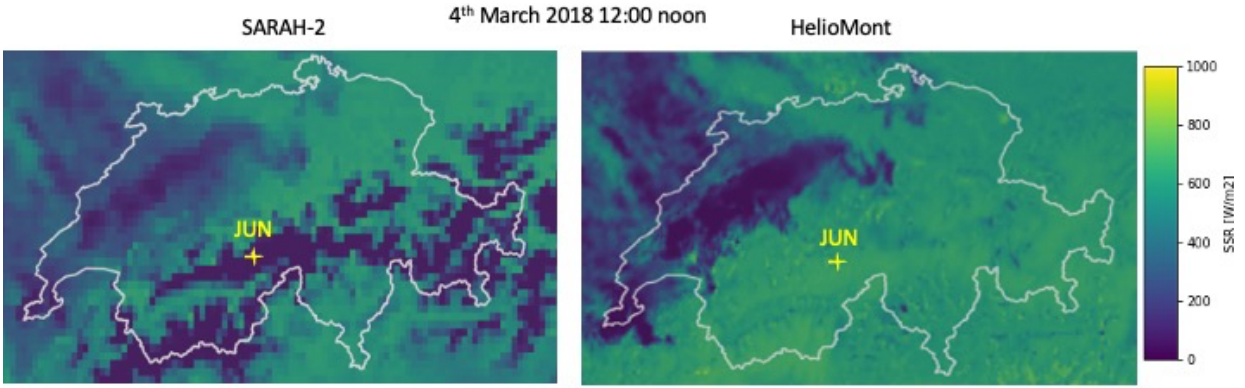

**Figure A2.** SSR maps of Switzerland from SARAH-2 and HelioMont for the exemplary case of 3 March 2018 at 12:00 noon. SARAH-2 misses out on SSR in the Alpine region, possibly due to misinterpretation of snow cover as clouds, see also Figure A3. JUN indicates the Alpine station at Jungfraujoch as an example of an affected site.

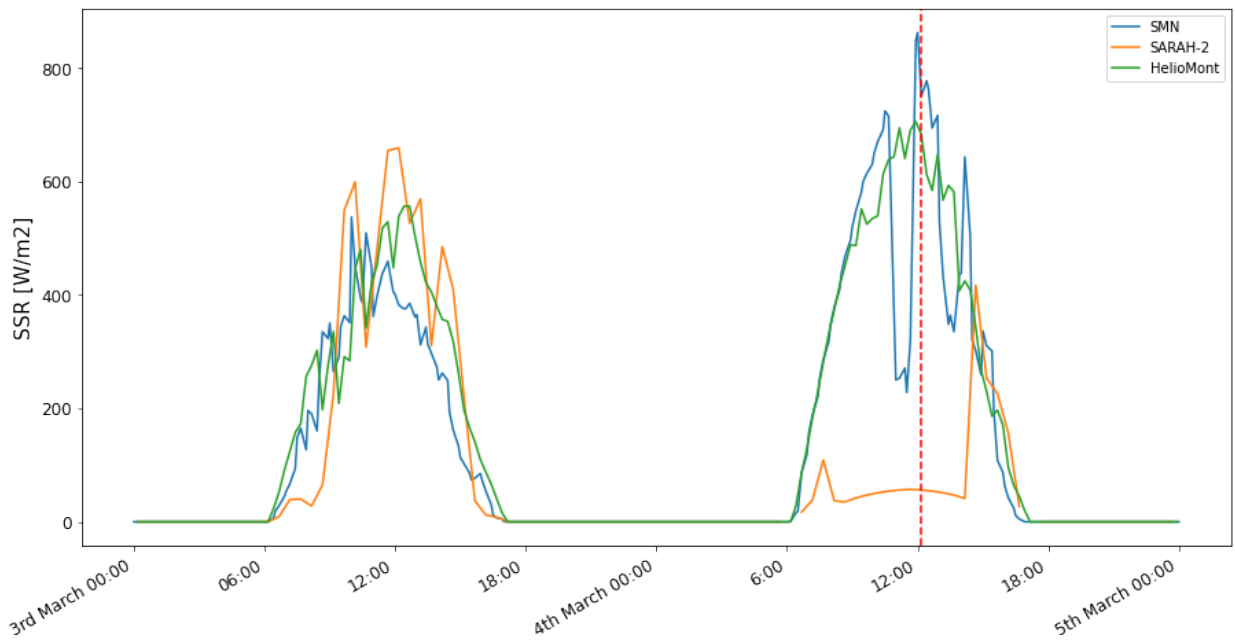

**Figure A3.** Exemplary SSR time series for two days in March 2018 at SMN station Jungfraujoch (JUN) to illustrate the SSR biases at higher-altitude sites. The time of SSR underestimation by SARAH-2 on 3 March 2018 at 12 a.m. shown in Figure A2 is indicated by the dashed line.